\documentclass[aps,pra,10pt,twocolumn]{revtex4-2}
\usepackage{graphicx}
\usepackage{amsmath,amsthm,amssymb,dsfont}
\usepackage{mdframed}
\usepackage{orcidlink}
\usepackage[dvipsnames]{xcolor}
\usepackage[normalem]{ulem}
\usepackage[switch]{lineno}

\usepackage{hyperref}

\begin{document}
\title{Enhanced quantum parameter estimation based on the Hardy paradox}

\author{Ming Ji}
\email{physmji@gmail.com}
\author{Yuxiang Yang}
\email{yuxiang@cs.hku.hk}
\affiliation{QICI Quantum Information and Computation Initiative, School of Computing and Data Science, The University of Hong Kong, Pokfulam Road, Hong Kong}
\author{Holger F. Hofmann}
\email{hofmann@hiroshima-u.ac.jp}
\affiliation{Graduate School of Advanced Science and Engineering, Hiroshima University, Kagamiyama 1-3-1, Higashi Hiroshima 739-8530, Japan}

\begin{abstract}
Statistical paradoxes such as the Hardy paradox and the enhancement of phase estimation via post-selection both draw upon the same non-classical features of quantum statistics described by non-positive quasi-probabilities. In this paper, we introduce a post-selected quantum metrology scenario where the initial state, the dynamics associated with the phase shift, and the post-selection are all inspired by the Hardy paradox. Specifically, we identify an anomalous weak value that is characteristic of both the Hardy paradox and the potential enhancement of sensitivity by the post-selection. We find that the efficiency of the enhancement is reduced when the expectation value associated with the anomalous weak value is different from the inverse of this value. We conclude that the relation between enhanced phase estimation and the Hardy paradox requires a detailed understanding of the relation between weak values and expectation values.
\end{abstract}

\maketitle

\textit{Introduction}---The non-classicality of quantum theory can be illustrated by a number of different aspects. Whether there exists a single fundamental concept that captures the essence of non-classicality remains an open question. If the answer is yes, we expect that quantum technologies that provide a quantum advantage should always draw upon this fundamental concept as a critical resource. 

One field where non-classical correlations are expected to provide a practical advantage is quantum metrology~\cite{Giovannetti2004Quantum,Giovannetti2006QUantum,Degen2017quantum}, where the goal is to estimate a classical parameter characterizing the statistical changes observed in a quantum system. The optimal precision of this estimate, also known as the quantum Cram$\mathrm{\acute{e}}$r-Rao bound, is quantified by the quantum Fisher information (QFI), which is bounded by the uncertainty $4\Delta S^2=4(\langle\hat{S}^2\rangle-\langle\hat{S}\rangle^2)$, where $\hat{S}$ is the generator of the original dynamics~\cite{HELSTROM1967minimum}. We remark that the QFI bound is tight for pure states. However, this limitation does not apply when the estimate is based on the conditional probabilities of only a fraction of outcomes in post-selected quantum metrology. It has been shown that the QFI of post-selected metrology can even exceed the limit set by the maximal uncertainty $4\Delta S^2$ permitted by the eigenvalues of $\hat{S}$ if the post-selection is described by negative quasi-probabilities of the Kirkwood-Dirac type~\cite{Arvidsson-Shukur2020KD,Lupu2022}. Negative quasi-probabilities can thus be viewed as a resource in post-selected quantum metrology.

Negative quasi-probabilities also provide a natural description of a wide variety of quantum paradoxes
~\cite{Spekkens2008NegCont,Ferrie2011quasi,Hofmann2024statistical,Arvidsson2024properties}. This suggests that quantum paradoxes can tell us something about the way in which negative quasi-probabilities and their associated anomalous weak values~\cite{Aharonov1988Weak} provide the enhancement of post-selected metrology beyond the quantum Cram$\mathrm{\acute{e}}$r-Rao bound the QFI bound $4\Delta S^2$ of the initial state. 
The paradoxical correlations between different quantum measurements can be 
captured by the concept of contextuality~\cite{kochen1967problem,Budroni2022ContextualityReview}, which has already been identified as a quantum resource in quantum computation~\cite{Raussendorf2001quan} and quantum cryptography~\cite{Bechmann2000Cryptography}. Contextuality is often associated with negative quasi-probabilities that can be represented by negative weak values of projection operators~\cite{Tollaksen2007pre,Williams2008weak,hofmann2023sequential,Ji2024tracing}.
Ref.~\cite{Arvidsson-Shukur2020KD} indicates that a negative quasi-probability in the generator's post-selected statistics is needed for the enhancement of QFI beyond the maximal uncertainty $4\Delta S^2$ permitted by the eigenvalues of $\hat{S}$. We might therefore expect that the corresponding resource can be represented by an anomalous weak value of the generator related to the dynamics of the phase estimation. It has been shown that the negative weak values associated with quantum paradoxes describe half-periodic unitary transformations, relating them directly to a corresponding parameter estimation problem where the inequality violation appears as an anomalous weak value of the generator \cite{Hofmann2015}. It is therefore possible to identify, from quantum paradoxes, the generator of the dynamics for which an anomalous QFI enhancement is possible. This method can be applied to any paradox related to contextuality~\cite{Leifer2005pre,hance2023contextuality,Ji2024quantitative}, providing a starting point for the conversion of contextuality into enhanced quantum metrology.

In this paper, we will take a look at the well-known Hardy paradox, since it is both related to quantum non-locality and to contextuality ~\cite{Hardy1992Quantum,Hardy1993Nonlocality,Cabello2013Simple,ji2023nonclassical}. We design a quantum parameter estimation experiment in which both the generator in the unitary transformation and the post-selection are associated with the measurements and outcomes involved in the Hardy paradox. The post-selection is designed so that the post-selection probability depends directly on the observation of the Hardy paradox, thus identifying the violation of classical expectations in the paradox with one of the essential resources in post-selected parameter estimation. The advantage of using the Hardy paradox is that it has one free parameter, given by the overlap of two local eigenstates. We explore the relation between this free parameter and the enhanced QFI, identifying the efficiency of the enhancement based on the comparison between the bound set by the post-selection probability and the actual QFI observed. The equations show that, in addition to the post-selection probability, the difference between the expectation value of the generator and the inverse of the anomalous weak value associated with the paradoxical outcome needs to be suppressed as well. The latter insight may be an important step towards a better understanding of the physics described by anomalous weak values.

\textit{Contextuality in the Hardy paradox}---We start by formulating the Hardy paradox for a pair of identical two-level systems $1$ and $2$ as an example of quantum contextuality. Each system possesses two properties $\hat{W}_i$ and $\hat{F}_i\ (i=1,2)$, where $\hat{W}=|b\rangle\langle b|-|a\rangle\langle a|$ and $\hat{F}=|1\rangle\langle1|-|0\rangle\langle0|$ with $\langle a|b\rangle=\langle0|1\rangle=0$. The eigenstates of the different operators are related to each other by superpositions:
\begin{align}
    |a\rangle&=\langle 0|a\rangle \;|0\rangle \;+ \langle 1 |a \rangle\;|1\rangle,\nonumber\\
    |b\rangle&=\langle 1|a\rangle^* |0\rangle- \langle 0|a\rangle^*|1\rangle,
\end{align}
where $|\langle0|a\rangle|^2+|\langle1|a\rangle|^2=1$. 
According to the Kochen-Specker theorem, a context is defined by a set of commuting operators that can be measured jointly
~\cite{kochen1967problem,Budroni2022ContextualityReview}. In the Hardy paradox, these are the four combinations of local operators given by $\{\hat{F}_1, \hat{F}_2\}$, $\{\hat{F}_1, \hat{W}_2 \}$, $\{\hat{W}_1, \hat{F}_2 \}$ and $\{\hat{W}_1,\hat{W}_2\}$. Each measurement occurs in two of the four contexts, making it possible to relate the statistics of different contexts by their shared measurements. A non-contextual hidden variable theory would assign the same outcome to each local measurement, independent of the measurement performed on the other system. The Hardy paradox demonstrates that the statistics observed separately in the four different contexts cannot be explained by assigning non-contextual outcomes to each of the local measurements. 

We start our discussion by considering the outcome $|a,a\rangle$ of the $\{\hat{W}_1,\hat{W}_2\}$ measurement. If the local outcomes are independent of context, we can consider the outcomes of $\hat{F}_1$ and $\hat{F}_2$ associated with the outcome $|a,a\rangle$. Either both outcomes are $1$, or at least one of the two outcomes is $0$. We can then divide the probability $P(a,a)$ into four parts taken from the four different contexts and derive the non-contextual inequality,
\begin{equation}
\label{eq:inequality}
    P(a,a)\leq P(a,0)+P(0,a)+P(1,1).
\end{equation}
This inequality defines the bound that applies if the outcomes $(a,a)$, $(a,0)$, $(0,a)$ and $(1,1)$ are related to each other by a simple `either/or' logic for the values of $\hat{F}_i$. A violation of the inequality shows that this kind of logic does not apply in quantum mechanics. In the Hardy paradox, this violation is obtained by requiring that the right-hand side of Eq.~(\ref{eq:inequality}) is zero. In quantum mechanics, this is represented by three orthogonality relations~\cite{ji2023nonclassical},
\begin{align}
\label{eq:orthogonality}
    &\langle\phi_0|a,0\rangle=0,\nonumber\\
    &\langle\phi_0|0,a\rangle=0,\nonumber\\
    &\langle\phi_0|1,1\rangle=0.
\end{align}
In the four-dimensional Hilbert space of the two two-level systems, these orthogonality relations uniquely define the quantum state
\begin{equation}
\label{eq:initialstate}
    |\phi_0\rangle=\frac{-\langle 1|a\rangle^* |0,0\rangle+\langle 0 | a \rangle^*|0,1\rangle+\langle 0 | a \rangle^*|1,0\rangle}{\sqrt{1+|\langle 0|a \rangle|^2}}.
\end{equation}
Contrary to non-contextual logic, the only state satisfying the three conditions in Eq.~(\ref{eq:orthogonality}) has a non-vanishing probability for the outcome $|a,a\rangle$ given by
\begin{align}
\label{eq:probabilityaa}
    P(a,a|\phi_0)
          &=|\langle 0|a\rangle|^4\frac{1-|\langle 0|a\rangle|^2}{1+|\langle 0|a\rangle|^2}.
\end{align}
In the widely discussed case of $|\langle 0|a\rangle|^2=1/2$, this probability is equal to $1/12$~\cite{Fra18}. Here, we will use the probability $P(a,a|\phi_0)$ as a quantitative measure of contextuality, making use of the full range of possible values of $|\langle 0|a\rangle|^2$, from zero to one. 

How do quantum superpositions produce such a paradoxical result? In the $\{\hat{F}_1,\hat{F}_2\}$ basis used in Eq.~(\ref{eq:initialstate}), the inner product of $|\phi_0\rangle$ and $|a,a\rangle$ is described by destructive interferences between $|0,0\rangle$ and $\{|0,1\rangle,|1,0\rangle\}$ defined by the conditions $P(a,0)=0$ and $P(0,a)=0$. However, the destructive interference in $P(a,a)$ is not fully balanced because it involves two positive terms and only one negative term. Destructive interference still ensures that the probability $P(a,a)$ is rather low, but the mismatch between the two interfering amplitudes leaves a non-vanishing probability that contradicts the non-contextual assignment of eigenvalues to $\hat{F}_1$ and $\hat{F}_2$. A quantitative measure of the destructive interference effect can be obtained by applying a phase flip operation described by the operator
 \begin{align}
\label{eq:Sdefinition}
   \hat{S} &= \hat{F}_1 \otimes \hat{F}_2.
\end{align}
The application of this operator to $|\phi_0\rangle$ aligns the phases of all three components for maximal constructive interference in $|a,a\rangle$~\cite{Hofmann2015}. The ratio between the amplitudes before and after the operation are given by an anomalous weak value,
 \begin{align}
\label{eq:anomalousWV}
   \frac{\langle a,a |\hat{S}|\phi_0 \rangle}{\langle a,a|\phi_0\rangle} &= - 3.
\end{align}
This anomalous weak value is a typical characteristic of the Hardy paradox that can be used to link it with other quantum phenomena. Here, we consider the relation with post-selected metrology, where the relevance of negative doubly extended Kirkwood-Dirac quasi-probabilities has been noted~\cite{Arvidsson-Shukur2020KD}. The anomalous weak value in Eq.~(\ref{eq:anomalousWV}) is an indicator that post-selection might enhance the ability to sense unitary phase shifts generated by $\hat{S}$.

\textit{A post-selected parameter estimation experiment based on the Hardy paradox}---We will now construct a parameter estimation scenario based on the initial state $|\phi_0\rangle$ and a unitary transformation generated by $\hat{S}$. The parameterized state reads
\begin{align}
\label{eq:theta}
|\phi(\theta)\rangle &= \exp(-i \theta \hat{S}) |\phi_0 \rangle.
\end{align}
In general, the phase sensitivity is given in terms of the uncertainty of an estimator $\hat{M}$, where 
\begin{align}
    \label{eq:estimate}
    \frac{1}{\delta \theta^2} &= \frac{\left(\frac{d}{d \theta} \langle \hat{M} \rangle \right)^2}{\Delta M^2}.
\end{align}
For optimized estimators $\hat{M}$, when the number of repetition $N$ of the estimation experiment satisfies $N\gg1$, the phase sensitivity achieves the quantum Cramér-Rao bound given by the quantum Fisher information (QFI) of
\begin{align}
    \label{eq:QFI}
    I_0 &= 4 \Delta S^2,
\end{align}
where $\Delta S^2 = 1-\langle \hat{S} \rangle^2$ is the uncertainty of the generator $\hat{S}$ in the states $| \phi(\theta) \rangle$. 

Without post-selection, the QFI only depends on the uncertainty of the generator, and this uncertainty is bounded by the difference between the extremal eigenstates of the generator.
However, it has been shown that post-selection can result in an enhancement of QFI  to values greater than $4 \Delta S^2$~\cite{Arvidsson-Shukur2020KD} in the post-selected subspace:
Instead of selecting an estimator $\hat{M}$ for the complete Hilbert space, one first performs a binary POVM $\{\hat I-\hat\Pi,\hat\Pi\}$ and selects only the outcomes $\hat\Pi$ for the phase estimation. This may result in an enhanced QFI if the parameter dependence is concentrated within the conditional probabilities observed in the subspace selected by $\hat\Pi$. Since the QFI is a measure of information per outcome, the omission of the outcomes $\hat I-\hat\Pi$ will result in an increase of QFI unless the post-selection actively discards useful information on the parameter.
Effectively, a low post-selection probability can be traded for an enhanced conditional QFI. In general, the enhancement of the QFI is bounded by~\cite{Combes2014limits} 
\begin{align}
\label{eq:limit}
I_{\mathrm{select}} \leq \frac{4\Delta S^2}{P(\Pi|\phi_0)},
\end{align}
where $P(\Pi|\phi_0)$ is the post-selection probability. However, post-selected metrology is also limited by the dimensionality of the Hilbert space that remains after the post-selection, since removing too many dimensions through the post-selection would result in the loss of ``space" available for phase encoding, thereby limiting the phase sensitivity. It is therefore desirable to post-select a $(d-1)$-dimensional subspace of Hilbert space, removing only a single high probability outcome from the final measurement. This creates a problem for the application of paradoxical quantum statistics to post-selected metrology. The weak value in Eq.~(\ref{eq:anomalousWV}) is defined by the post-selection of the paradoxical outcome $|a,a \rangle$, not by the elimination of a high probability outcome.
In order to establish a relation between the Hardy paradox and post-selected metrology, we must therefore identify an outcome with a particularly high probability that is nevertheless related to the low probability $P(a,a|\phi_0)$ that violates the inequality given in Eq.~(\ref{eq:inequality}). We can do so by referring to the anomalous weak value in Eq.~(\ref{eq:anomalousWV}). The square of this weak value gives the ratio between the probability of $| a,a \rangle$ and the probability of $\hat{S} |a,a \rangle$. If we define the post-selected Hilbert space in terms of the projection operator
\begin{align}
    \label{eq:postselection}
    \hat{\Pi} = \hat{1} - \hat{S} |a,a \rangle \langle a,a| \hat{S},
\end{align}
the post-selection probability $P(\Pi|\phi_0)=\langle \hat{\Pi} \rangle$ is related to the probability $P(a,a|\phi_0)$ of the Hardy paradox by
\begin{align}
    \label{eq:PiProbability}
    P(\Pi|\phi_0) = 1 - 9 P(a,a|\phi_0),
\end{align}
where the factor of nine is the squared anomalous weak value given in Eq.~(\ref{eq:anomalousWV}). By defining the post-selection strategy according to Eq.~(\ref{eq:postselection}), we have ensured that the QFI enhancement originates from the Hardy paradox and not from other non-classical correlations described by the state $|\phi_0\rangle$. The larger $P(a,a|\phi_0)$, the lower the post-selection probability and the higher the possibility of an enhanced post-selected QFI.
Therefore, high probabilities $P(a,a|\phi_0)$ can result in greater enhancements of the post-selected QFI beyond the uncertainty limit of $4\Delta S^2$, establishing a direct link between the violation of the inequality in Eq.~(\ref{eq:inequality}) and the achievement of enhanced QFI in post-selected metrology. We would like to stress that this is the only efficient post-selection strategy that satisfies this requirement. Since the goal of the present analysis is the evaluation of the metrological advantage associated with the inequality violation of Hardy's paradox, it is not possible to optimize the choice according to any additional figures of merit. It is certainly possible to obtain metrological advantages by other post-selection strategies, and these may be related to other quantum paradoxes. Here, we focus on the question whether the Hardy paradox is sufficient for the observation of enhanced QFI in a post-selection scenario that depends directly on the characteristic inequality violation of the paradox.

A low post-selection probability is a necessary condition for the enhancement of quantum metrology, but the precise value of the post-selected QFI includes additional terms that describe a reduction of $I_{\mathrm{select}}$ below the limit in Eq.~(\ref{eq:limit}). Following the derivation in~\cite{Arvidsson-Shukur2020KD}, the precise value of the post-selected QFI is
\begin{align} 
\label{eq:enhancedQFI}
I_{\mathrm{select}}=&\frac{4 \langle \phi_0 | \hat{S} \hat{\Pi} \hat{S} | \phi_0 \rangle}{\langle \phi_0 | \hat{\Pi} | \phi_0 \rangle} - \frac{4 |\langle \phi_0 | \hat{\Pi} \hat{S} | \phi_0 \rangle|^2}{(\langle \phi_0 | \hat{\Pi} | \phi_0 \rangle)^2}.
\end{align}
Using Eqs.~(\ref{eq:postselection}) and (\ref{eq:PiProbability}) together with the anomalous weak value in Eq.~(\ref{eq:anomalousWV}), the post-selected QFI can be expressed as a function of the post-selection probability $P(\Pi|\phi_0)$, the uncertainty $\Delta S^2=1-\langle \hat{S}\rangle^2$, and the expectation value $\langle\hat{S} \rangle$,
\begin{align} 
\label{eq:efficiency}
I_{\mathrm{select}}=\frac{4 \Delta S^2}{P(\Pi|\phi_0)} \left(1 - \frac{(1-P(\Pi|\phi_0))}{P(\Pi|\phi_0)} \; \frac{(\langle \hat{S} \rangle + 1/3)^2}{\Delta S^2}\right).
\end{align}
When given in this form, it is found that the bound of Eq.~(\ref{eq:limit}) is only achieved when $\langle \hat{S} \rangle= -1/3$. Eq.~(\ref{eq:enhancedQFI}) indicates that this value is equal to the inverse of the anomalous weak value in Eq.~(\ref{eq:anomalousWV}). The efficiency of the enhancement achieved thus depends on the closeness of the expectation value of $\hat{S}$ to the inverse of the anomalous weak value that characterizes the paradox. Since the expectation value of $\hat{S}$ for the state $|\phi_0 \rangle$ is given by 
\begin{align}
\label{eq:expectationS}
    \langle \hat{S} \rangle = \frac{1-3 |\langle 0 | a \rangle|^2}{1+|\langle 0 | a \rangle|^2},
\end{align}
the bound of Eq.~(\ref{eq:limit}) is only achieved for $|\langle0|a\rangle|^2=1/2$, with probabilities of $P(a,a|\phi_0)=1/12$ and $P(\Pi|\phi_0)=1/4$. The post-selection probability then results in a four-fold enhancement of the QFI over the uncertainty limit of $4\Delta S^2=4-4/9$, easily exceeding even the upper limit of $4$ for states with a maximal uncertainty of $\Delta S^2=1$. 

\textit{Contextuality and the enhancement of post-selected QFI}---We have shown that the statistics of the Hardy paradox can be converted into a quantum metrology scenario where the violation of classical expectations serve as a necessary resource for the post-selected enhancement of QFI. Both the choice of the generator $\hat S$ and the post-selection $\hat \Pi$ are motivated by the algebra of the paradox. We can now take a closer look at the quantitative relation between the enhancement of post-selected QFI and the statistics of the Hardy paradox.

\begin{figure}
    \centering
    \includegraphics[width=\linewidth]{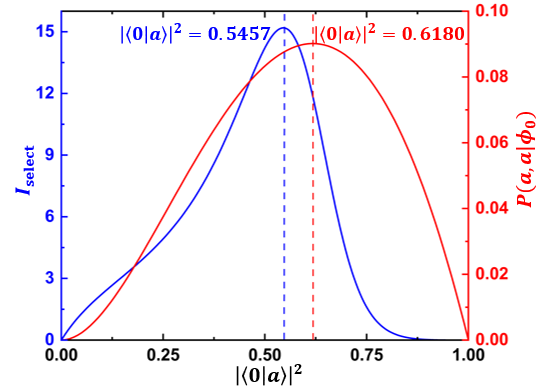}
    \caption{Post-selected QFI $I_\mathrm{select}$ and contextual probability $P(a,a|\phi_0)$ as a function of the free parameter $|\langle 0 | a \rangle|^2$ characterizing the Hardy paradox. The dotted lines indicate the location of the maximal values.}
    \label{fig:QFIprobability}
\end{figure}

As mentioned above, the probability $P(a,a|\phi_0)$ is a direct measure of the amount of contextuality, since it describes the violation of the inequality~(\ref{eq:inequality}) for the conditions that define $| \phi_0 \rangle$. We can interpret this quantitative measure of contextuality as a non-classical resource that can also be applied to enhanced quantum metrology. As indicated by Eq.~(\ref{eq:PiProbability}), increasing the value of $P(a,a|\phi_0)$ would decrease the value of the post-selection probability $P(\Pi|\phi_0)$. Given that the enhancement of QFI is bounded by the inverse post-selection probability as shown in Eq.~(\ref{eq:limit}), it is obvious that the definition of post-selection in Eq.~(\ref{eq:postselection}) activates quantum contextuality as a resource for enhanced parameter estimation. However, Eq.~(\ref{eq:efficiency}) shows that the efficient use of this resource depends on the closeness of the expectation value of $\hat{S}$ to the inverse of the anomalous weak value. It is therefore useful to consider the post-selected QFI as a function of the free parameter $|\langle0|a\rangle|^2$. Combining Eqs.~(\ref{eq:probabilityaa}), (\ref{eq:PiProbability}), (\ref{eq:efficiency}), (\ref{eq:expectationS}) and the uncertainty $\Delta S^2=1-\langle \hat{S}\rangle^2$, we have
\begin{equation}
    \label{eq:QFI_innerproduct}
    I_\mathrm{select}=\frac{32|\langle0|a\rangle|^2\left(1-|\langle0|a\rangle|^2\right)^3}{\left(1+|\langle0|a\rangle|^2-9|\langle0|a\rangle|^4+9|\langle0|a\rangle|^6\right)^2}.
\end{equation}
Fig.~\ref{fig:QFIprobability} shows the QFI $I_\mathrm{select}$ obtained after post-selection, as given by Eq.~(\ref{eq:QFI_innerproduct}). An enhancement above the limit of $I_0=4$ set by the maximal uncertainty of the generator $\hat{S}$ is observed for a wide range of values of $|\langle 0 | a \rangle|^2$, with a peak value of $I_\mathrm{select}=15.2067$ at $|\langle0|a\rangle|^2=0.5457$. This value is almost four time higher than the maximal QFI that can be achieved without post-selection, demonstrating the success of the post-selection strategy. To investigate the relation with contextuality, Fig.~\ref{fig:QFIprobability} also shows the contextual probability $P(a,a|\phi_0)$, as given by Eq.~(\ref{eq:probabilityaa}). As expected, the maximum of QFI $I_\mathrm{select}$ and the maximum of contextual probability $P(a,a|\phi_0)$ are very close to each other, with the maximal value of $I_\mathrm{select}$ at $|\langle0|a\rangle|^2=0.5457$ and the maximal value of $P(a,a|\phi_0)$ at $|\langle0|a\rangle|^2=0.6180$. It may also be worth noting that the probability $P(a,a|\phi_0)$ at the maximum of $I_\mathrm{select}$ is already at $0.0875$, very close to its maximum value of $0.0902$. This can be taken as strong evidence that the successful enhancement of QFI by post-selection observed close to the peak of $I_\mathrm{select}$ draws upon the contextual probability $P(a,a|\phi_0)$ as a resource. 

However, the picture that emerges for large values of $|\langle 0 | a \rangle|^2$ is quite different. Even though contextuality persists, the post-selected QFI drops to nearly zero, indicating a complete failure to access the available quantum contextuality. As noted before, this happens because the expectation value of $\hat{S}$ is very different from $-1/3$. Between $|\langle 0 | a \rangle|^2=1/2$ and $|\langle 0 | a \rangle|^2=1$, this expectation value drops from $-1/3$ to $-1$, increasing the deviation from the ideal value of $-1/3$ while simultaneously reducing the initial QFI $I_0$. To separate these two effects, it is useful to introduce the conversion efficiency $\eta$ that quantifies how much of the initial QFI is successfully concentrated into the post-selected outcome $\hat \Pi$. The conversion efficiency can be expressed as the ratio of $I_{\mathrm{select}}$ and the limit given by Eq.(\ref{eq:limit}),
\begin{equation}
    \label{eq:conversionefficiency}
    \eta=\frac{P(\Pi|\phi_0)I_{\mathrm{select}}}{4\Delta S^2}.
\end{equation}
This efficiency is given by the term in brackets in Eq.~(\ref{eq:efficiency}) and is equal to one minus a term proportional to the squared deviation of $\langle \hat{S} \rangle$ from $-1/3$. The reduction of efficiency is enhanced when the uncertainty $\Delta S^2$ and the post-selection probability $P(\Pi|\phi_0)$ are low. 

\begin{figure}
    \centering
    \includegraphics[width=\linewidth]{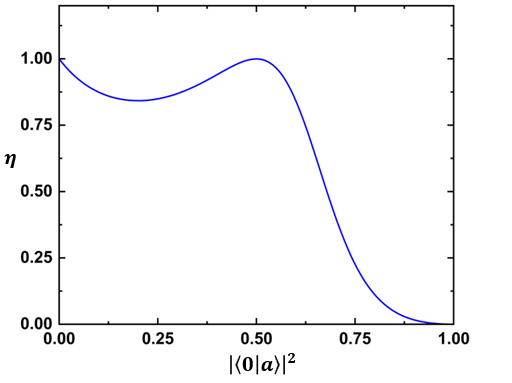}
    \caption{Conversion efficiency $\eta$ as a function of the free parameter $|\langle 0 | a \rangle|^2$. The conversion efficiency drops sharply just below $|\langle 0 | a \rangle|^2=0.75$, indicating that contextuality cannot be accessed for enhanced parameter estimation when $|\langle 0 | a \rangle|^2$ is above that value. }
    \label{fig:efficiency}
\end{figure}

Fig.~\ref{fig:efficiency} shows the dependence of efficiency $\eta$ on the free parameter $|\langle0|a\rangle|^2$. Although a small drop in efficiency is observed below the maximum of one at $|\langle 0 | a \rangle|^2=1/2$, this drop is not very deep since the efficiency is back to one at $|\langle 0 | a \rangle|^2=0$. It might be worth noting that this high efficiency is associated with a post-selection probability of $P(\Pi|\phi_0)=1$, indicating only that the small reductions of post-selection probability for $|\langle 0 | a \rangle|^2\ll 1$ are directly converted into QFI increases. The same is not true for the drop observed above the maximum at $|\langle 0 | a \rangle|^2=1/2$. Here, the decrease in efficiency continues all the way to a value of zero at $|\langle 0 | a \rangle|^2=1$. As a point of reference, an efficiency of $1/2$ is obtained for $|\langle 0 | a \rangle|^2=0.6787$. All efficiencies at lower values of  $|\langle 0 | a \rangle|^2$ are greater than $1/2$, all at higher values are lower than $1/2$.

The analysis above focused on the total QFI achieved and on the efficiency of enhancement relative to the maximal enhancement allowed by the post-selection probability. While this analysis identifies the role of contextuality as a resource, it does not indicate whether an enhancement was actually achieved or not. For this purpose, we should consider the ratio $I_\mathrm{select}/I_\mathrm{0}$ of enhanced QFI and initial QFI before post-selection. When expressed as a function of the free parameter $|\langle 0 | a \rangle|^2$, this ratio is given by 
\begin{equation}
    \label{eq:QFIratio}
    \frac{I_\mathrm{select}}{I_0}=\left(\frac{1-|\langle0|a\rangle|^4}{1+|\langle0|a\rangle|^2-9|\langle0|a\rangle|^4+9|\langle0|a\rangle|^6}\right)^2.
\end{equation}

\begin{figure}
    \centering
    \includegraphics[width=\linewidth]{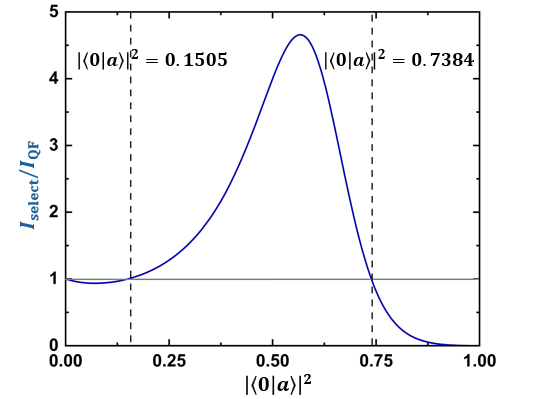}
    \caption{Ratio between QFI $I_\mathrm{select}$ enhanced by post-selection $\hat{\Pi}$ and QFI $I_0$ without post-selection. The value of this ratio measures the enhancement of the QFI induced by the post-selection. Post-selection enhances QFI for $I_\mathrm{select}/I_\mathrm{0}>1$, observed in the range of  $[0.1505,0.7384]$ for the free parameter $|\langle0|a\rangle|^2$.}
    \label{fig:QFIratio}
\end{figure}

The enhancement is shown in Fig.~\ref{fig:QFIratio}. Note that the enhancement drops below one at both high and low values of $|\langle 0 | a \rangle|^2$, indicating that the reduction of efficiency can remove the advantages of post-selection completely. A successful enhancement of $I_\mathrm{select}/I_0>1$ is observed when the free parameter $|\langle0|a\rangle|^2$ is between $0.1505$ and $0.7384$. Outside this range, the efficiency $\eta$ is too low to achieve any enhancement. Below $0.1505$, the high post-selection probability associated with low contextual probabilities $P(a,a|\phi_0)$ provides only very little possible enhancement to start with, causing a drop below $1$ even for the small dip in efficiency shown in Fig.~\ref{fig:efficiency}. Above $0.7384$, the efficiency is below $0.2616$, explaining the lack of enhancement despite the availability of sufficient contextual probability. The maximum of contextual probability is obtained at $|\langle0|a\rangle|^2=0.6180$, which is in the upper part of this range. As noted above, an enhancement of $4$ is obtained at $|\langle0|a\rangle|^2=1/2$. We can expect to find the maximal enhancement between these two values, marking the maximal available resource and the maximal efficiency, respectively. Indeed, the maximum of $I_\mathrm{select}/I_0$ is found at $|\langle0|a\rangle|^2=0.5671$, very close to the average of these two values. The maximal enhancement achieved by optimizing the balance between contextuality as a resource and efficiency of conversion is $4.6647$ at an efficiency of $\eta=0.9349$ and contextual probability of $0.0888$. 

Here, we focused on linking the post-selection probability $P(\Pi|\phi_0)$ with the contextual probability $P(a,a|\phi_0)$ based on the anomalous weak value in Eq.~(\ref{eq:anomalousWV}). Our results show that this is highly efficient as long as the expectation value of $\hat{S}$ is close to the inverse weak value of $-1/3$. It may be interesting to consider the physics behind this limitation. Post-selected parameter estimation achieves enhancements of QFI by removing outcomes that carry no useful information about the parameter to be estimated. The same sensitivity is then concentrated in the remaining outcomes, resulting in the maximal possible enhancement by the inverse of the post-selection probability. The reduction of efficiency can then be explained by information about the parameter that was lost when outcomes were discarded. We can conjecture that this information loss depends on the difference between the expectation value and the inverse weak value. With respect to the post-selection process, the inverse weak value is equal to the weak value of $\hat{S}$ for the discarded outcome $\hat{S}| a,a \rangle$. The efficiency is therefore limited by the inadvertent loss of information associated with the discarded state, as expressed by the difference between the expectation value of $\hat{S}$ and the weak value for the discarded outcome $\hat{S}| a,a \rangle$. 

\textit{Conclusions}---
Anomalous weak values can explain both quantum paradoxes and quantum advantages such as the enhancement of QFI in a post-selected subspace. It should therefore be possible to learn more about the physics behind both by investigating the relation between the two. In this work, we have introduced a systematic approach towards the application of contextuality observed in the Hardy paradox to enhanced parameter estimation. We successfully devised a scenario that makes use of the non-zero contextual probability $P(a,a|\phi_0)$ that characterizes Hardy's paradox as a resource for enhanced parameter estimation by identifying a post-selection strategy where the post-selection probability $P(\Pi|\phi_0)$ decreases as the contextual probability $P(a,a|\phi_0)$ increases. The detailed analysis of this scenario has shown that the use of this resource is limited by the efficiency $\eta$. This efficiency is a characteristic of the mechanism of enhanced parameter estimation, where optimal enhancement requires that the discarded outcomes do not carry any information about the parameter of interest. In the post-selection scenario devised here, this is not automatically achieved. However, it is worth noting that the ideal case of an efficiency of one is achieved for the case of $|\langle 0 | a \rangle|^2=1/2$, where the operators $\hat{F}_i$ and $\hat{W}_i$ can be represented by Pauli operators. Very low efficiencies indicating a high amount of information in the discarded outcome are only observed for high values of $|\langle 0 | a \rangle|^2$, with an efficiency of more than $1/2$ observed for all values below $0.6787$. In conclusion, we have shown that the contextual probability that characterizes Hardy's paradox can be used to enhance parameter estimation if the parameter estimation scenario is developed according to the conditions of the paradox. The detailed analysis of the scenario has shown that enhanced parameter estimation is further limited by an inadvertent loss of information carried by the discarded states, and this loss of information is not related to contextuality. The non-zero probability associated with Hardy's paradox serves as a resource for post-selected metrology, but its conversion into an enhancement of QFI requires further conditions that can be expressed in terms of the relation between expectation values and weak values in the quantum statistics of the system. Our analysis of the relation between post-selected metrology and Hardy's paradox thus reveals statistical details that go beyond the role of negative quasi-probabilities investigated in previous work \cite{Arvidsson-Shukur2020KD,Lupu2022}.

\section*{Acknowledgment}
M. Ji and Y. Yang acknowledge support from the National Science Foundation of China via the Excellent Young Scientists Fund (Hong Kong and Macau) Project 12322516, the National Natural Science Foundation of China (NSFC)/Research Grants Council (RGC) Joint Research Scheme via Project N\_HKU7107/24, and the Hong Kong Research Grant Council (RGC) through the General Research Fund (GRF) grant 17302724. H. F. Hofmann acknowledges support from ERATO, Japan Science and Technology Agency (JPMJER2402).

\bibliographystyle{unsrturl}
\bibliography{ref.bib}

\end{document}